\documentclass[runningheads]{llncs}

\usepackage[numbers]{natbib}

\usepackage[utf8]{inputenc}
\usepackage{relsize}
\usepackage{underscore}
\usepackage{graphicx}
\usepackage{hyperref}
\usepackage[firstpage]{draftwatermark}
\usepackage{enumitem} 
\usepackage{booktabs}
\usepackage{tabulary}
\usepackage{standalone}
\usepackage[many]{tcolorbox}
\tcbset{frame empty,colback=yellow!20,boxsep=0mm,}

\SetWatermarkAngle{0}
\SetWatermarkText{\raisebox{11.5cm}{%
  \hspace{0.1cm}%
  \href{https://doi.org/10.5281/zenodo.6521395}{\includegraphics{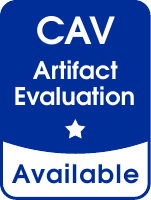}}%
  \hspace{9cm}%
  \includegraphics{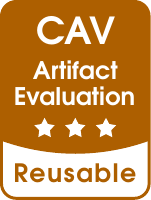}%
}}

\def\Inf{\ensuremath{\mathsf{Inf}}}
\def\Fin{\ensuremath{\mathsf{Fin}}}
\newcommand{\LTLF}{\mathsf{F}} 
\newcommand{\LTLG}{\mathsf{G}} 
\newcommand{\LTLX}{\mathsf{X}} 
\newcommand\Cpp{C\nolinebreak\hspace{-.05em}\raisebox{.4ex}{\relsize{-3}{\textbf{+}}}\nolinebreak\hspace{-.10em}\raisebox{.4ex}{\relsize{-3}{\textbf{+}}}}

\definecolor{lime}{HTML}{A6CE39}
\DeclareRobustCommand{\orcidicon}{
	\hspace{-2.5mm}
	\begin{tikzpicture}[baseline={(0,-0.12)}]
	\draw[lime, fill=lime] (0,0)
	circle [radius=0.16]
	node[white] (ID) {{\fontfamily{qag}\selectfont \tiny ID}};
	\draw[white, fill=white] (-0.0625,0.095)
	circle [radius=0.007];
	\end{tikzpicture}
	\hspace{-2.5mm}
      }
\def\orcidID#1{\href{https://orcid.org/#1}{\smash{\orcidicon}}}

\title{From Spot 2.0 to Spot 2.10: What's New?}
\authorrunning{A. Duret-Lutz, et al.}
\author{
Alexandre~Duret-Lutz\inst{1}\orcidID{0000-0002-6623-2512}
\and Etienne~Renault\inst{1}\orcidID{0000-0001-9013-4413}
\and Maximilien~Colange\inst{2}\orcidID{0000-0003-4769-3302}
\and Florian~Renkin\inst{1}\orcidID{0000-0002-5066-1726}
\and Alexandre~Gbaguidi~Aisse\inst{2}
\and Philipp~Schlehuber-Caissier\inst{1}\orcidID{0000-0002-6611-9659}
\and Thomas~Medioni\inst{2}
\and Antoine~Martin\inst{1}\orcidID{0000-0002-3263-7669}
\and Jérôme~Dubois\inst{1}
\and Clément~Gillard\inst{2}
\and Henrich~Lauko\inst{2}\orcidID{0000-0002-5422-5884}
}

\institute{LRDE, EPITA, Le Kremlin-Bicêtre, France\\
  \email{\{adl,\,renault,\,frenkin,\,philipp,\\amartin,\,jdubois\}@lrde.epita.fr}
  \and
  Previously at LRDE.
}
\begin{document}

\maketitle

\vspace*{-1ex}

\begin{abstract}
  Spot is a \Cpp17 library for LTL and $\omega$-automata manipulation,
  with command-line utilities, and Python bindings.  This paper
  summarizes its evolution over the past six years, since the
  release of Spot 2.0, which was the first version to support
  $\omega$-automata with arbitrary acceptance conditions, and the last
  version presented at a conference.  Since then, Spot has been
  extended with several features such as acceptance transformations,
  alternating automata, games, LTL synthesis, and more.  We also shed
  some lights on the data-structure used to store automata.
  \par
  \textbf{Artifact:} \url{https://zenodo.org/record/6521395}.
\end{abstract}

\section{Availability, Purpose, and Evolution}

Spot is a library for LTL and $\omega$-automata manipulation,
distributed under a GPLv3 license.  Its source code is available from
\url{https://spot.lrde.epita.fr/}.  We provide packages for some Linux
distributions like Debian and Fedora, but other packages can also be
found for Conda-Forge~\cite{conda.15.misc} (for Linux \& Darwin), Arch
Linux, FreeBSD...

Spot can be used via three interfaces: a \Cpp17 library, a set of
command-line tools that give easy access to many features of the
library, and Python bindings, that makes prototyping and interactive
work very attractive.  Our web site now contains many examples of how
to perform some tasks using these three interfaces, and we have a
public mailing list for questions.

In our last tool paper~\cite{duret.16.atva2}, Spot~2.0 had just
converted from being a library for working on Transition-based
Generalized Büchi Automata and had become a library supporting
$\omega$-automata with arbitrary
Emerson-Lei~\cite{emerson.87.scp,safra.89.stoc} acceptance conditions,
as enabled by the development of the HOA format~\cite{babiak.15.cav}.

%

In the HOA format, transitions can carry multiple colors, and
acceptance conditions are expressed as a positive Boolean formulas
over atoms like $\Fin(i)$ or $\Inf(i)$ that tell if a color should be
seen finitely or infinitely often for a run to be accepting.
Table~\ref{tab:acc} gives some examples.

\begin{table}[tb]
  \caption{Acceptance formulas corresponding to classical names.
    \label{tab:acc}}
  \centering
  \setlength{\tabcolsep}{8pt}
  \def\rowfill{}
  \begin{tabular}{ll}
    \toprule
    Büchi & $\Inf(0)$ \rowfill \\
    generalized Büchi & $\bigwedge_i\Inf(i)$ \rowfill \\
    \Fin-less~\cite{bloemen.19.sttt} & any positive formula of $\Inf(...)$\rowfill \rowfill \\
    co-Büchi & $\Fin(0)$ \rowfill \\
    generalized co-Büchi & $\bigvee_i\Fin(i)$ \rowfill \\
    Rabin & $\bigvee_i \left(\Fin(2i)\land\Inf(2i+1)\right)$ \rowfill \\
    generalized Rabin~\cite{kretinsky.12.cav} & $\bigvee_i(
                        \Fin(i)\land\bigwedge_{j\in J_i}\Inf(j))$ \rowfill \\
    Streett & $\bigwedge_i \left(\Inf(2i)\lor\Fin(2i+1)\right)$ \rowfill \\
    parity min even & $\Inf(0) \lor (\Fin(1) \land (\Inf(2) \lor (\Fin(3) \land \ldots)))$ \rowfill \\
    parity min odd & $\Fin(0) \land (\Inf(1) \lor (\Fin(2) \land (\Inf(3) \lor \ldots)))$ \rowfill \\
    parity max even & $(((\Inf(0) \land \Fin(1))\lor \Inf(2))\land \Fin(3))\lor \ldots$ \\
    parity max odd & $(((\Fin(0) \lor \Inf(1))\land \Fin(2))\lor \Inf(3))\land \ldots$ \\
    \bottomrule
  \end{tabular}
\end{table}

While Spot 2.0 was able to read automata with arbitrary acceptance
conditions, not all of its algorithms were able to support such a
generality.  For instance testing an automaton for emptiness or
finding an accepting word, would only work on automata with
``\Fin-less'' acceptance conditions.  For other conditions, Spot 2.0
would rely on a procedure called \texttt{remove_fin()} to convert
automata with arbitrary acceptance conditions into ``\Fin-less''
acceptance conditions~\cite{bloemen.19.sttt}.  This was ultimately
fixed by developing a generic emptiness check~\cite{baier.19.atva}.
Additionally the support for arbitrary acceptance conditions has
allowed us to implement many useful algorithms; the most recent being
the Alternating Cycle
Decomposition~\cite{casares.21.icalp,casares.22.tacas} a powerful data
structure with many applications (conversion to parity acceptance,
degeneralization, typeness
checks...)\footnote{\url{https://spot.lrde.epita.fr/ipynb/zlktree.html}}.


There have been 56 releases of Spot since version 2.0, but only 10 of
these are major releases.  Releases are numbered $2.x.y$ where $y$ is
updated for minor upgrades that mostly fix bugs, and $x$ is updated
for major release that add new features.  (The leading $2$ would be
incremented in case of a serious redesign of the API.)
Table~\ref{tab:milestones} summarizes the highlights of the various
releases in chronological order.  Not appearing in this list are many
micro-optimizations and usability improvements that Spot has
accumulated over the years.

\begin{table}[tbp]
  \caption{Milestones in the history of Spot. \label{tab:milestones}}
\begin{tabulary}{\textwidth}{cccJ}
2004 & 0.x & \Cpp03 & Prehistory of the project. \cite{duret.04.mascots} \\
2012 & 0.9 &       & Support for some PSL operators. \\
2013 & 1.0 &       & Command-line tools, mostly focused on LTL/PSL input~\cite{duret.13.atva}.  Includes
\texttt{ltlcross}, a clone of LBTT~\cite{tauriainen.99}.  Python bindings. \\
     & 1.1 &       & Automatic detection of stutter-invariant formulas.~\cite{michaud.15.spin}\\
     & 1.2 &       & SAT-based minimization~\cite{baarir.14.forte,baarir.15.lpar}. \texttt{ltlcross} and the new \texttt{dstar2tgba} can read Rabin and Streett automata produced by \texttt{ltl2dstar}~\cite{klein.07.ciaa}. \\
2016 & 2.0 & \Cpp11 & Rewrite of the LTL formulas representation.  Rewrite of the automaton class to allow arbitrary acceptance.  Support for the HOA format.  More command-line tools, now that automata can be exchanged with other tools.~\cite{duret.16.atva2}  New determinization procedure. \\
     & 2.1 &       & Conversion to generalized Streett or Rabin.  Small usability improvements all around (like better support for CSV files). \\
     & 2.2 &       & LTLf$\to$LTL conversion~\cite{degiacomo.13.ijcai}.  Faster simulation-based reduction of deterministic automata. \\
2017 & 2.3 &       & Initial support for alternating automata and alternation removal.  400\% faster emptiness check.  Incremental SAT-based minimization.  Classification in the temporal hierarchy of Manna \& Pnueli~\cite{manna.90.podc}. \\
     & 2.4 & \Cpp14 & New command-line tools: \texttt{autcross} to check and compare automata transformations, \texttt{genaut} to generate families of automata.  Dualization of automata.  Conversion from Rabin to Büchi~\cite{krishnan.94.isaac} updated to support transition-based input.  Relabeling of LTL formulas with large Boolean subformulas to speedup their translation.  \\
2018 & 2.5 &       & New command-line tool \texttt{ltlsynt} for synthesis of AIGER circuits from LTL specifications.~\cite{michaud.18.synt}   Conversions to co-Büchi~\cite{boker.11.fossacs}.  Utilities for converting between parity acceptance conditions.  Detection of stutter-invariant \emph{states}.  Determinization optimized.\\
     & 2.6 &       & Compile-time option to support more than 32 colors.  Specialized translation for
formulas of the type $\LTLG\LTLF(\varphi)$ if $\varphi$ is a guarantee.  New translation mode to output
automata with unconstrained acceptance condition.  Semi-deterministic complementation~\cite{blahoudek.16.atva}.  Faster detection of obligation properties.  Online LTL translator replaced by a new web application (see Figure~\ref{fig:webapp}).\\
     & 2.7 &       & LAR-based paritization in \texttt{ltlsynt}.  Generic emptiness check~\cite{baier.19.atva}.  Detection of liveness properties~\cite{alpern.87.dc}. \\
2019 & 2.8 &       & Accepting run extraction for arbitrary acceptance.  Introduction of an ``\texttt{output_aborter}'' to abort constructions that are too large.  Support for SVA's delay syntax, and \texttt{first_match} operator~\cite{systemverilog.18.std}.  Minimization of parity acceptance~\cite{carton.99.ita}.\\
2020 & 2.9 &       & Better paritization, partial degeneralization, and acceptance simplifications~\cite{renkin.20.atva}.  Weak and strong variants of $\LTLX$.  Xor product of automata, used while translating
formulas to automata with unconstrained acceptance. \\
2021 & 2.10 & \Cpp17 & \texttt{ltlsynt} overhauled~\cite{renkin.21.synt}.  Support for games and Mealy machines.  Mealy machines simplifications.  Multiple encodings from Mealy machine to AIGER.  Experimental \texttt{twacube} class for parallel algorithms.  Support for transition-based Büchi.   Zielonka Trees and Alternating Cycle Decomposition~\cite{casares.21.icalp,casares.22.tacas} \\
\end{tabulary}
\end{table}

\section{Use-cases of Spot, and Related Tools}

As it is a library, there are many ways to use Spot.  We are mostly
aware of such uses via citations\footnote{Our previous tool
  paper~\cite{duret.16.atva2} has over 250 citations according to
  \href{https://scholar.google.ca/scholar?oi=bibs&cites=3741341698957703284}{Google
    scholar}}.  Historical and frequent uses-cases are to use Spot for
translating LTL formulas to automata (Winners of the sequential LTL
and parallel LTL tracks of RERS'19 challenge~\cite{jasper.19.rers}
both used Spot to translate the properties into automata, many
competitors on the Model Checking Contest~\cite{kordon.21.mcc} also
use Spot this way), or to use it as a research/development toolbox,
since it provides helper tools for generation of random
formulas/automata, verification of LTL-to-automata translation,
simplifications, syntax conversions, etc.  Nowadays, the algorithms
for $\omega$-automata implemented in Spot are often used as baseline
for studying better
algorithms~\cite[e.g.,][]{loding.19.atva,kretinsky.21.ai,halvena.22.cav,doveri.22.cav},
but we also see some new applications built on top of
$\omega$-automata algorithms from
Spot~\cite[e.g.,][]{brotherston.12.aplas,bruyere.22.arxiv}.

The projects that have the largest intersections of features with Spot
seem to be GOAL~\cite{tsai.13.cav} and Owl~\cite{kretinsky.atva.18}.
These are two Java-based frameworks that deal with similar objects and
provide a range of algorithms.
Owl and Spot share a similar and traditional Unix view of the
command-line experience, where multiple commands are expected to be
chained with pipes, and they both communicate smoothly via the HOA
format~\cite{babiak.15.cav}.  GOAL is centered on a graphical
interface in which the user can edit automata, and apply algorithms
listed in menu entries.  Using GOAL from the command-line is possible
by writing short scripts in a custom language.

As far as interfacing goes, the most important feature of Spot is
probably that it exposes its algorithms and data structures in Python.
Beside being usable as a glue language between various tools, this
allows us (1) to leverage Python's ecosystem and (2) to quickly
prototype new algorithms in Python.

\section{Automata Representation}

In this section and the next three, we focuses on how the storage of
automata evolved to support alternation, games, and Mealy machines.

The main automaton class of Spot is called \texttt{twa_graph} and
inherits from the \texttt{twa} class.  The letters \texttt{twa} stand
for \emph{Transition-based $\omega$-Automaton}.

The class \texttt{twa} implements an abstract interface that allows
on-the-fly exploration of an automaton similar to what had been
present in Spot from the start: essentially, one can query the initial
state, and query the transitions leaving any known state.  In
particular, before exploring the state-space of a \texttt{twa}, it is
unknown how many states are reachable.  Various subclasses of
\texttt{twa} are provided in Spot, for instance to represent the
state-space of Promela or Divine models~\cite{duret.16.atva2}.  Users
may create subclasses, for instance to create a Kripke
structure on-the-fly.\footnote{As demonstrated by \url{https://spot.lrde.epita.fr/tut51.html}}

\begin{figure}[tb]
  \includegraphics[width=\textwidth]{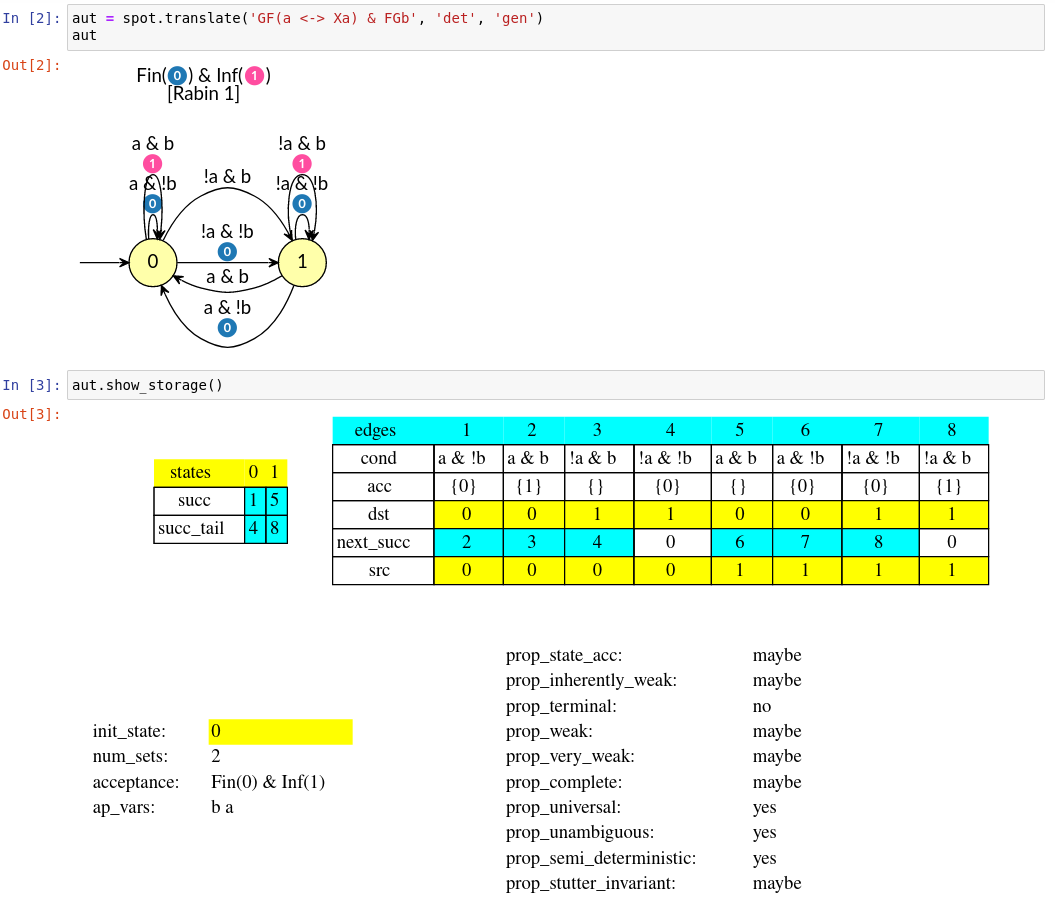}
  \caption{Internal representation of a \texttt{twa_graph} as two vectors.\label{fig:storage}}
\end{figure}

The class \texttt{twa_graph}, introduced in Spot 2.0, implements an
explicit, graph-based, representation of an automaton, in which states
and edges are designated by integers.  This makes for a much simpler
interface\footnote{Contrast on-the-fly and explicit APIs at
  \url{https://spot.lrde.epita.fr/tut50.html}.} and usually simplifies
the data structures used in algorithms (since states and edges can
be used as indices in arrays).  The data structure is best
illustrated by using the \texttt{show_storage()} method of the Python
bindings, as shown by Figure~\ref{fig:storage}.
A \texttt{twa_graph} is stored as two \Cpp{} vectors: a
vector of states, and a vector of edges.  For each state, the first
vector stores two edge numbers: \texttt{succ} is the first outgoing
edge, and \texttt{succ_tail} is the last one.  These number are
indices into the edge vector, which stores five pieces of information
per edge.  Four of them are related to the identity of the edge:
\texttt{src}, \texttt{dst}, \texttt{cond}, \texttt{acc} are
respectively the source, destination, guard, and color sets of the
edge.  The remaining field, \texttt{next_succ} gives the next outgoing
edge, effectively creating a linked list of all edges leaving a given
state.  There is no edge 0: this value is used as terminator for such
lists.  Outgoing edges of the same state are not necessarily adjacent
in that structure.  When a new edge is added to the automaton, it is
simply appended to the edge vector, and the \texttt{succ_tail} field
of the state is used to update the previous end of the list.

To iterate over successors of state 1 in \Cpp{} or Python, one can ignore the above linked list implementation and write one of the following loops:\\
\begin{minipage}{.495\textwidth}
\begin{tcolorbox}[left=2mm]
\begin{verbatim}
for (auto& e: aut->out(1))
  // use e.cond, e.acc, e.dst
\end{verbatim}
\end{tcolorbox}
\end{minipage}
\begin{minipage}{.495\textwidth}
\begin{tcolorbox}[left=2mm]
\begin{verbatim}
for e in aut.out(1):
    # use e.cond, e.acc, e.dst
\end{verbatim}
\end{tcolorbox}
\end{minipage}\\
The \texttt{twa_graph::out} methods simply returns a lightweight
temporary object which can be iterated upon using iterators that will
follow the linked list.  Then the object \texttt{e} is effectively
a reference to a column of the edge vector.

As seen on Figure~\ref{fig:storage}, the automaton additionally stores
an initial state (Spot only supports a single initial state), a number
of colors (\texttt{num_sets}), an acceptance condition, a list of
atomic propositions (Spot only supports alphabets of the form
$2^{\mathit{AP}}$), and 10 fields storing structural properties of the
automaton.

These property fields have only three possible values: they default to
\emph{maybe}, but can be set to \emph{no} or \emph{yes} by algorithms
that work on the automaton.  They can also be read and written in the
HOA format.  For instance if \texttt{prop_universal} is set to
\emph{yes}, it means that automaton does not have any existential
choice (a.k.a. non-determinism).  Spot's \emph{is_deterministic()}
algorithm can return in constant time if \texttt{prop_universal} is
known, otherwise it will inspect the automaton and set that property
before returning, so that the next call to \emph{is_deterministic()}
will be instantaneous.  Some algorithms know how to take
advantage of any hint they get from those properties: for instance the
\texttt{product()} of two automata is optimized to use fewer colors
when one of the arguments is known to be weak (i.e., in an SCC all
transitions have the same colors).

Note that algorithms that modify an automaton in place have to
remember to update those properties.  This has caused a couple of bugs
over the years.

\section{Introduction of Alternating Automata}

Support for alternating $\omega$-automata, as defined in the HOA
format, was added to Spot in version 2.3 without introducing a new
class.  Rather, the \texttt{twa_graph} class was extended to support
alternation in such a way that existing algorithms would not require
any modification to continue working on automata without universal
branching.  This was done by reserving the sign bit of the destination
state number of each transition to signal universal branching.

\begin{figure}[tb]
  \includegraphics[width=\textwidth]{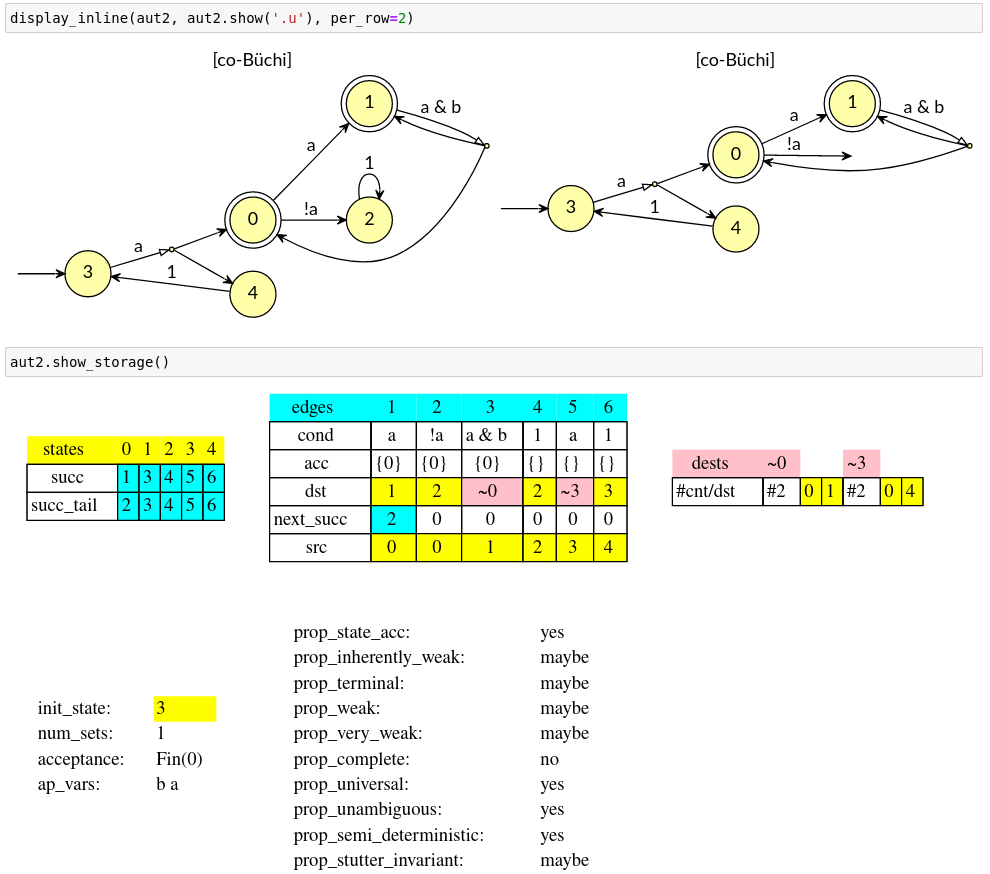}
  \vspace*{-2em}
  \caption{Internal representation of alternating automata.\label{fig:alt}}
\end{figure}

Figure~\ref{fig:alt} shows an example of Alternating automaton
(top-left) with co-Büchi acceptance.  In many works on alternating
automata, it is conventional to not represent accepting sinks, and
instead have transition without destination.  The top-right picture
shows that Spot has a rendering option to hide accepting sinks.

The bottom of the figure shows that the automaton has
\texttt{prop_state_acc} set, which means that the automaton is meant
to be interpreted as using state-based acceptance.  Colors are still
stored on edges internally, but all edges leaving a state have the
same colors.  Seeing that the condition is co-Büchi ($\Fin(0)$), the
display code automatically switched to the convention of using
double-circles for rejecting states.

Destinations with the sign bit set are called \emph{universal
  destination groups} and appear as pink in the figure.  There are two
groups here: \texttt{\textasciitilde{}0} and
\texttt{\textasciitilde{}3}.  The complement of these numbers can be
used as indices in the \texttt{dests} vector, that actually store the
destination groups.  At the given index, one can read the size $n$ of
the destination group, followed by the state number of the $n$
destinations.

Algorithms that work on alternating automata need to be able to iterate
over all destinations of an edge.  The process of checking the sign bit
of the destination to decide if its a group, and to iterate on that group
is hidden by the \texttt{univ_dests()} method:\\
\begin{minipage}{.54\textwidth}
\begin{tcolorbox}[left=1pt]
\begin{verbatim}
for(auto& e: aut->out(1)) {
 // use e.cond, e.acc, e.src
 for(unsigned d:aut->univ_dests(e))
  // use d
}
\end{verbatim}
\end{tcolorbox}
\end{minipage}
\begin{minipage}{.460\textwidth}
\begin{tcolorbox}[left=1pt]
\begin{verbatim}
for e in aut.out(1):
  # use e.cond, e.acc, e.src
  for d in aut.univ_dests(e):
     # use d

\end{verbatim}
\end{tcolorbox}
\end{minipage}\\
Note that this code works on non-universal branches as well: if \texttt{e.dst}
is unsigned, \texttt{univ_dests(e)} will simply iterate on that unique value.

Spot has two alternation removal procedures.  One is an on-the-fly
implementation of the Breakpoint construction~\cite{miyano.84.tcs}
which transforms an $n$-state alternating Büchi automaton into a
non-alternating Büchi automaton with at most $3^n$ states.  For very
weak alternating automata, it is known that a powerset-based procedure
can produce a transition-based generalized Büchi automaton with $2^n$
states~\cite{gastin.01.cav}; in fact that algorithm even works on
\emph{ordered} automata~\cite{boker.10.icalp}, i.e., alternating
automata where the only rejecting cycles are self-loops.  The second
alternation removal procedure of Spot is a mix between these two
procedures but does not work on the fly: it takes a \emph{weak}
automaton as input, and uses the break-point construction on
rejectings SCCs that have more than one state, and uses the powerset
construction for other SCCs.

\section{Extending Automata via Named Properties}

Spot's automata have a mechanism to attach arbitrary data to
automata, called \emph{named properties}.  (This is similar to
the notion of attributes in the R language.)   An object can be attached
to the automaton with:
\begin{tcolorbox}[left=1pt]
\begin{verbatim}
  aut->set_named_prop("property-name", new mytype(...));
\end{verbatim}
\end{tcolorbox}
\noindent and later retrieved with:
\begin{tcolorbox}[left=1pt]
\begin{verbatim}
  mytype* data = aut->get_named_prop<mytype>("property-name");
\end{verbatim}
\end{tcolorbox}
Ensuring that \texttt{mytype} is the correct type for the retrieved
property is the programmer's responsability.

Spot has grown a list of many such properties over
time.\footnote{\url{https://spot.lrde.epita.fr/concepts.html\#named-properties}}
For instance \texttt{automaton-name} stores a string that would be
displayed as the name of the automaton.  The \texttt{highlight-edges}
and \texttt{highlight-states} properties can be used to color edges
and states.  The \texttt{state-names} is a vector of strings that
gives a name to each state, etc.  While those examples are mostly
related to the graphical rendering of the automata, some algorithms
store useful byproducts as properties.  For instance the
\texttt{product()} algorithm will define a \texttt{product-states}
named property that store a vector of pairs of the original states.

These named properties are sometimes used to provide additional
semantics to the automaton, for instance to obtain a game or a Mealy
machine.

\section{Games, Mealy Machines, and LTL Synthesis}

The application of Spot to LTL synthesis was introduced in Spot~2.5 in
the form of the \texttt{ltlsynt} tool~\cite{michaud.18.synt}, but the
inner workings of this tool were progressively redesigned and publicly
exposed until version 2.10.

An automaton can now be turned into a game by attaching the
\texttt{state-player} property to
it.\footnote{\url{https://spot.lrde.epita.fr/tut40.html} illustrates
  how a game can be used to decide if a state simulates another one.}
Only two-player games are supported, so \texttt{state-player} should
be a \texttt{std::vector<bool>}.  Currently, Spot has solvers for
safety games and for games with \emph{parity max odd} acceptance, but
we plan to at least generalize the latter to any kind of parity
condition.  Once a game has been solved, it contains two new named
properties: \texttt{state-winner} (a \texttt{std::vector<bool>}
indexed by state numbers indicating the player winning in each state),
and \texttt{strategy} (a \texttt{std::vector<unsigned>} that gives for
each state the edge that its owner should follow to win).

\begin{figure}[tb]
  \includegraphics[width=\textwidth]{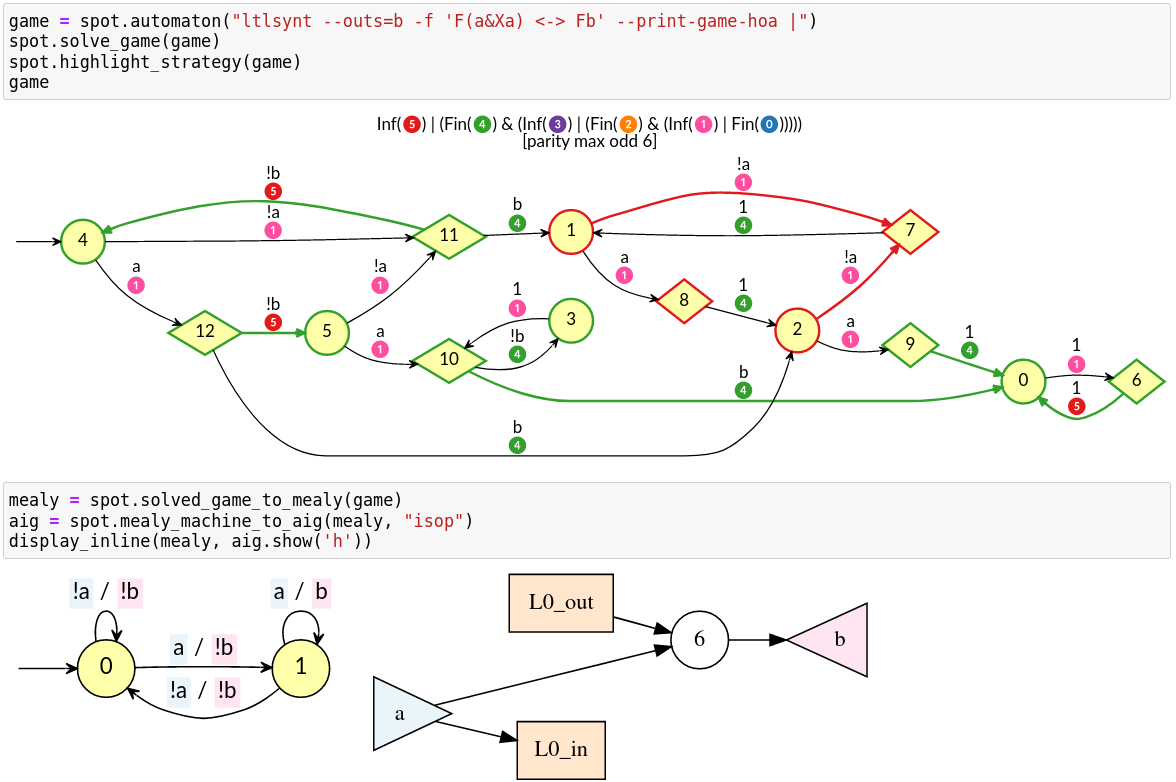}
  \caption{(top) Solving a game to display the strategy.  States with
    green borders are winning for player 1, who wants to satisfy the
    acceptance condition, by following the green arrows.  States with
    red color are winning for player 0, who wants to fail the
    acceptance condition, by following the red arrows. (bottom)
    Conversion of the winning strategy to a Mealy machine and then an
    AIGER circuit.\label{fig:game}}
\end{figure}
\begin{figure}[tb]
\centering
  \begin{minipage}[b]{.465\textwidth}
    \includegraphics[width=\textwidth]{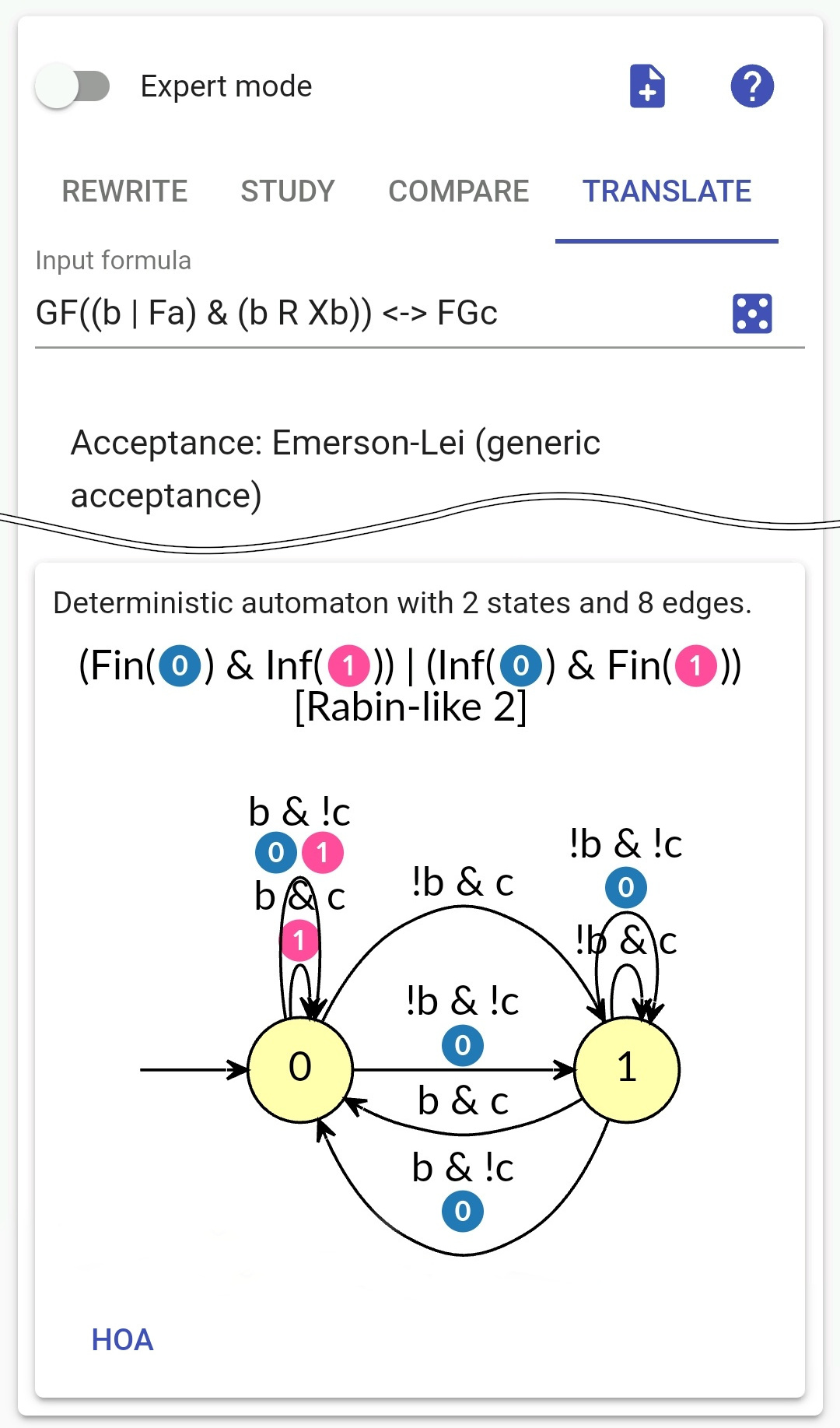}
  \end{minipage}
  \begin{minipage}[b]{.465\textwidth}
    \includegraphics[width=\textwidth]{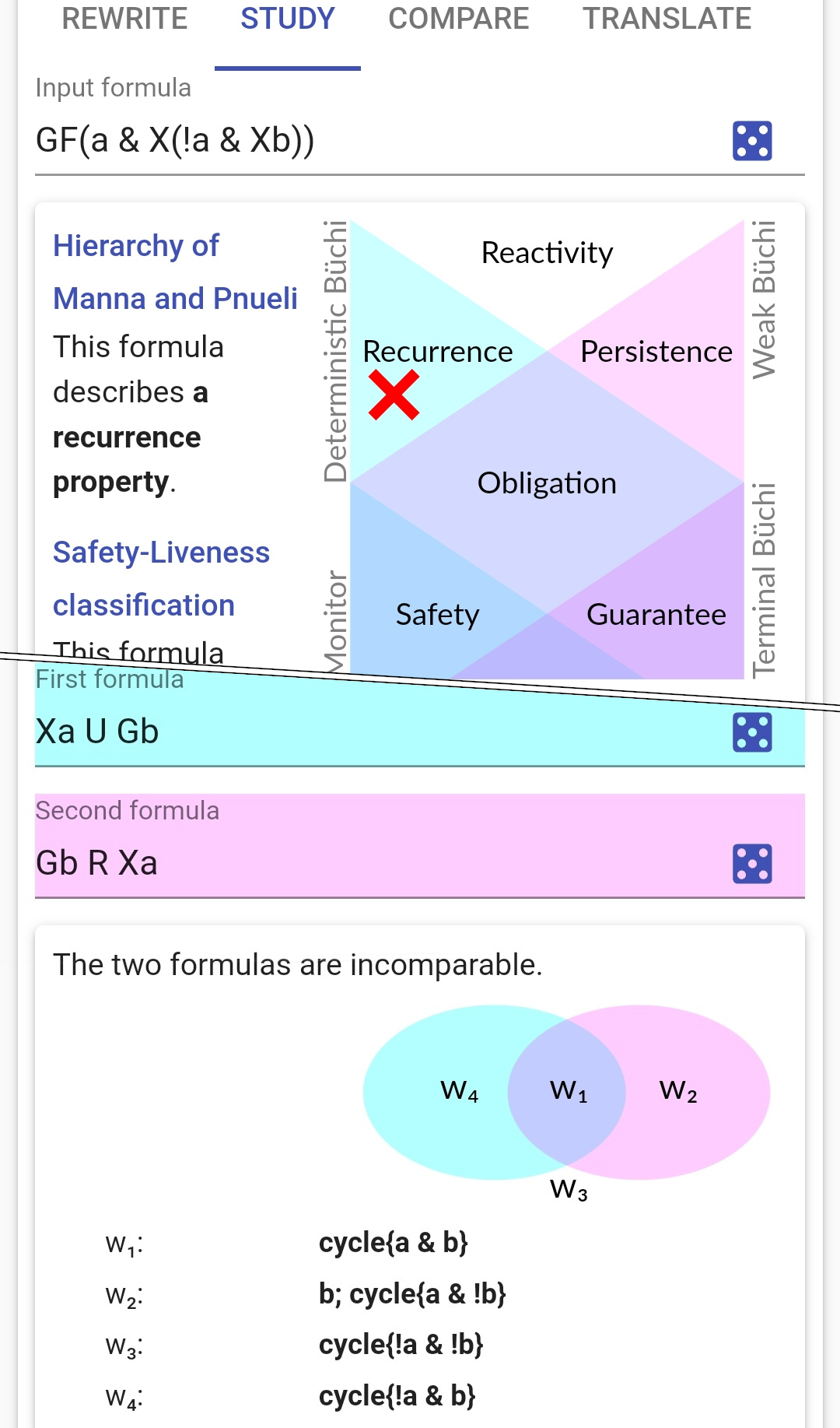}
  \end{minipage}
  \caption[A web application]{A web application, built on top of Spot.
    \url{https://spot.lrde.epita.fr/app/} \label{fig:webapp}}
\end{figure}

Figure~\ref{fig:game} shows an example of game generated by
\texttt{ltlsynt}, and how we can display the winning strategy once the
game is solved.  The winning strategy can be extracted and converted
into a Mealy machine, which is just an automaton that uses the
\texttt{synthesis-output} property to specify which atomic
propositions belong to the output.  Such a Mealy machine can then be
encoded into an AND-inverter graph, and saved into the AIGER
format~\cite{Biere-FMV-TR-11-2}.  Here L0 represents a latch,
i.e., one bit of memory, that stores the previous value of $a$ so
that the circuit can output $b$ if and only if $a$ is true in the
present and in the previous step.

\section{Online Application for LTL Formulas}\label{sec:webapp}

The Python ecosystem makes it easy to develop web interfaces for
convenient access to a subset of features of Spot.  For instance
Figure~\ref{fig:webapp} shows screenshots of a web application built
using a React frontend, and running Spot on the server.  It can
transform LTL formulas into automata, can display many properties of a
formula (membership to the Manna \& Pnueli
hierarchy~\cite{manna.90.podc}, Safety/Liveness
classification~\cite{alpern.87.dc}, Rabin and Streett
indices~\cite{carton.99.ita},
stutter-invariance~\cite{michaud.15.spin}), or simply compare two
formulas using a Venn diagram.

This application has been found to be useful for teaching about LTL
and its relation with automata, but is also a helpful research tool.

\section{Shortcomings and one Future Direction}

While Spot has been used for many applications, there are two
recurrent issues: they are related to the types used for some fields
of the edge vector (see Figures~\ref{fig:storage}--\ref{fig:alt}).  By
default, the set of colors that labels an edge (the \texttt{acc}
field) is stored as a 32-bit bit-vector, the transition label
(\texttt{cond}, a formula over $2^\mathit{AP}$), is stored as a BDD
identified by a unique 32-bit integer, and the other three fields
(\texttt{src}, \texttt{dst}, \texttt{next_succ}) are all 32-bit
integers.  One edge therefore takes 20 bytes.

While limiting the number of states to 32-bit integers has never
been a problem so far, the limit of 32 colors can be hit easily.
Spot~2.6 added a compile-time option to enlarge the number of
supported colors to any multiple of 32; this evidently has a memory
cost (and therefore also a runtime cost) as the \texttt{acc} field
will be larger for each edge.  However this constraint generally means
that all the algorithms we implement try to be ``color-efficient'', i.e.,
to not introduce useless colors.
For instance while the product of an automaton with $x$ colors and an
automaton with $y$ colors is usually an automaton with $x+y$ colors,
the \texttt{product()} implementation will output fewer colors in presence
of a \emph{weak} automaton.

The use of BDDs as edge labels causes another type of issues.  Spot
uses a customized version of the BuDDy library, with additional
functions, and several optimizations (more compact BDD nodes for
better cache friendliness, most operations have been rewritten to be
recursion-free).  However BuDDy is inherently not thread safe, because
of its global unicity table and caches.  This prevents us from doing
any kind of parallel processing on automata.  A long term plan is to
introduce a new class \texttt{twacube} that represent an automaton in
which edges are cubes (i.e., conjunctions of literals) represented
using two bit-vectors.  Such a class was experimentally introduced in
Spot 2.10 and is currently used in some parallel emptiness check
procedures~\cite{renault.16.sttt}.

\bibliographystyle{abbrvnat}
\bibliography{mc}
\end{document}